\documentclass[prd,11pt,a4paper,showpacs, nofootinbib, preprintnumbers,multicol]{revtex4}

\usepackage{latexsym}
\usepackage{epsfig}
\usepackage{amsfonts,amsmath}
\usepackage{amssymb}
\usepackage{graphicx}
\usepackage{bbm}
\usepackage{bm}
\usepackage{pxfonts}

\usepackage{color}
\definecolor{rosso}{cmyk}{0,1,1,0.4}
\definecolor{rossos}{cmyk}{0,1,1,0.55}
\definecolor{rossoc}{cmyk}{0,1,1,0.2}
\definecolor{blu}{cmyk}{1,1,0,0.3}
\definecolor{blus}{cmyk}{1,1,0,0.6}
\definecolor{bluc}{cmyk}{1,1,0,0.1}
\definecolor{verde}{cmyk}{0.92,0,0.59,0.25}
\definecolor{verdec}{cmyk}{0.92,0,0.59,0.15}
\definecolor{verdes}{cmyk}{0.92,0,0.59,0.4}

\makeatletter

\newcommand{\beq}{\begin{equation}}
\newcommand{\eeq}{\end{equation}}
\newcommand{\bea}{\begin{eqnarray}}
\newcommand{\eea}{\end{eqnarray}}
\newcommand{\ba}{\begin{array}}
\newcommand{\ea}{\end{array}}
\newcommand{\bi}{\begin{itemize}}
\newcommand{\ei}{\end{itemize}}
\newcommand{\bn}{\begin{enumerate}}
\newcommand{\en}{\end{enumerate}}
\newcommand{\bc}{\begin{center}}
\newcommand{\ec}{\end{center}}

\newcommand{\gsim}{\lower.7ex\hbox{$\;\stackrel{\textstyle>}{\sim}\;$}}
\newcommand{\lsim}{\lower.7ex\hbox{$\;\stackrel{\textstyle<}{\sim}\;$}}

\begin{document}


\title{\Large \color{red} \bf The Hypothesis of Superluminal Neutrinos:\\ comparing OPERA with other Data}

\author{Alessandro Drago$^{\color{blue}{\clubsuit }}$} 
\author{Isabella Masina$^{\color{blue}{\clubsuit} \color{blue}{\varheartsuit}}$}
\email{masina@fe.infn.it}
\author{Giuseppe Pagliara$^{\color{blue}{\clubsuit} }$}
\author{ Raffaele Tripiccione$^{\color{blue}{\clubsuit} }$}

\affiliation{
$^{\color{blue}{\clubsuit}}$ 
{Dip.~di Fisica dell'Universit\`a di Ferrara and INFN Sez.~di Ferrara, Via Saragat 1, I-44100 Ferrara, Italy}\\
$^{\color{blue}{\heartsuit}}$  
{CP$^\mathbf 3$-Origins and DIAS, SDU University, \\ Campusvej 55, DK-5230 Odense M, Denmark}\\ }

\setcounter{page}{1}

\vskip 1cm

\begin{abstract}
~~~\\ {\bf Abstract:}
The OPERA Collaboration reported evidence for muonic neutrinos traveling slightly faster than light in vacuum.
While waiting further checks from the experimental community, here we aim at exploring some theoretical consequences of the hypothesis
that muonic neutrinos are superluminal, considering in particular the tachyonic and the Coleman-Glashow cases. 
We show that a tachyonic interpretation is not only hardly reconciled with OPERA data on energy dependence, but that it clashes with neutrino production from pion
and with neutrino oscillations. A Coleman-Glashow superluminal neutrino beam would also have problems with pion decay kinematics for the OPERA setup; 
it could be easily reconciled with SN1987a data, but then it would be very problematic to account for neutrino oscillations.   
 \end{abstract}
\baselineskip=15pt
\maketitle



\section{Introduction}

Recently the OPERA Collaboration \cite{opera} reported an early arrival time for CNGS muon neutrinos with respect to the one expected 
assuming neutrinos to travel at the speed of light in vacuum $c$.
The relative difference of the velocity of the muon neutrinos $v$ with respect to light quoted by OPERA is:
\beq
\frac{v-c}{c}  = (2.48 \pm 0.28 (stat) \pm 0.30 (sys)) \times 10^{-5}\,\,,
\eeq
 for a mean energy of the neutrino beam of $17$ GeV.

Similar hints, but with much less significance, were also reported for muon neutrino beams produced at Fermilab. 
Dealing with energies peaked at $3$ GeV, the MINOS Collaboration \cite{Adamson:2007zzb} found in 2007 that
$(v-c)/c =(5.1 \pm 3.9)\times 10^{-5}$. In 1979, a bound on the relative velocity of the muon with respect to muon neutrinos 
(with energies from $30$ to $200$ GeV) was also extracted:
$|v-v_\mu|/v_\mu \le 4 \times 10^{-5} $ \cite{Kalbfleisch:1979rm}.

While urging the experimental community to further check and debate on these results, in particular the most recent ones,
it is worth to explore which theoretical consequences would follow from the hypothesis that {\it the muon neutrino is a superluminal particle}.

This clearly requires a deep modification of the Standard Model (SM) of particle physics, that assumes particles to be {\it subluminal}. 
The energy and momentum of a subluminal particle of mass $m$ and velocity $\vec v$ are: 
$E= m c^2/\sqrt{1-\frac{v^2}{c^2}}$, $\vec p= m \vec v/\sqrt{1-\frac{v^2}{c^2}}$. They are related by the dispersion relation 
$E^2=p^2 c^2+m^2 c^4$, where $p=|\vec p|$, and the deviation from the speed of light is:
\beq
\frac{c-v}{c}  = \frac{c}{c+v} \left(\frac{ m c^2}{E}\right)^2\,\,.
\label{deltavB}
\eeq

Here we focus on two possibilities to account for a superluminal particle: the tachyon \cite{Recami:1984xh}
and the Coleman-Glashow particle \cite{Coleman:1997xq, Coleman:1998ti}.

The usual expressions for energy and momentum can be extended to the region $v>c$ provided we substitute in the 
numerator $m\rightarrow i \tilde m$, where $\tilde m$ is a real number. For such a particle, the energy and momentum are thus
$E= \tilde m c^2/\sqrt{\frac{v^2}{c^2}-1}$, $\vec p= \tilde m \vec v/\sqrt{\frac{v^2}{c^2}-1}$,
and satisfy the dispersion relation 
\beq
E^2=p^2 c^2-{\tilde m}^2 c^4\,\,.
\eeq
A tachyonic particle of mass $\tilde m c^2$ and energy $E$ then travels faster than $c$ by an amount
\beq
\frac{v-c}{c}  = \frac{c}{c+v} \left(\frac{\tilde m c^2}{E}\right)^2\,\,.
\label{deltav}
\eeq 
The deviation from the speed of light for a tachyon is thus opposite with respect to the one of a subluminal particle with the same mass, eq. (\ref{deltavB}).
In both cases, for energies much larger than the mass, the particle speed $v$ approaches $c$.
We display in fig. \ref{fig-dvsv} the relative speed of the tachyon with respect to light as a function of the tachyon mass and for selected energy values. 

The proposal that the neutrino could be a tachyon dates back to 1985 \cite{Chodos:1984cy}. 
In the light of the recent results, it is worth to consider this hypothesis as a possible explanation of the OPERA data. 
As we are going to discuss in section \ref{secT}, this interpretation is very problematic for various reasons: not only the deviation
from $c$ would depend on energy, but it would not even be possible for a pion to produce a tachyon with the mass in the range required to fit
the OPERA data.

Another proposal to account for a superluminal particle has been suggested by Coleman and Glashow (CG) \cite{Coleman:1997xq, Coleman:1998ti}.
The idea is that the i-th particle has, in addition to its own mass $m_i$, its own maximum attainable velocity $c_i$, and obeys the standard dispersion relation:
\beq
E_i^2=p_i^2 c_i^2+ m_i^2 c_i^4 \,\,.
\eeq
The CG muon neutrino can indeed account for the OPERA data without any trouble associated with its production from pion, as we are going to discuss in section \ref{secCG}.
To explain the observation of neutrinos associated in time with SN1987a, it is however necessary to introduce another neutrino with speed practically equal to $c$.
This brings severe problems to neutrino oscillations, so that even the CG muon neutrino appears not to be a fully satisfactory explanation.  

We draw our conclusions is section \ref{concl}.

\section{Problems of a Tachyonic Interpretation}
\label{secT}

\subsection{Energy independence of the early arrival times}

If the neutrinos produced at CERN are tachyons with mass $\tilde m $, after having travelled a distance $L \approx 730$ km, 
their associated early arrival time is $\delta t = \frac{L}{c} \frac{v-c}{c}$, with $\frac{L}{c} \approx 2.4$ ms.
Consider two tachyonic neutrino beams of energy $E_{1}$ and $E_{2}$, with $E_{1}\le E_{2}$ for definiteness.
As follows from eq.(\ref{deltav}), the ratio of their early arrival times $\delta t_1$ and $\delta t_2$ has a simple energy scaling:
\beq 
\frac{\delta t_1}{\delta t_2}  = \left(\frac{E_{2}}{E_{1}}\right)^2 \,\,.
\eeq
The early arrival time of a tachyon neutrino beam is indeed smaller the larger is its energy. In particular, 
for $E_{2} \approx 3 E_{1}$, one expects $\delta t_2 \approx \delta t_1/9$. 
The difference of the arrival times is thus negative: $\delta t_2 -\delta t_1 \approx -  \delta t_1$.

Now, the OPERA Collaborations considers two sample neutrino beams with mean energy equal to $\bar E_1=13.9$ GeV and $\bar E_2=42.9$ GeV respectively
\footnote{If neutrinos were tachyons, the energy reconstruction of the OPERA Collaboration should be revisited. However, for tachyonic masses smaller
than GeV, such effect is negligible for the sake of the present considerations.}. The ratio of these energies is indeed close to $3$.
However, the experimental values of the associated early arrival times are respectively $\delta t_1=(53.1 \pm 18.8 (stat) \pm 7.4 (sys))$ ns and 
$\delta t_2=(67.1 \pm 18.2 (stat) \pm 7.4 (sys))$ ns. These data display no evidence of an energy dependence. 
OPERA quotes a value $\delta t_2 -\delta t_1 =(14.0\pm 26.2)$ ns for the difference of the arrival times $\delta t_2-\delta t_1$.
Far from being close to $-\delta t_1$ as expected for a tachyon, the latter value is even slightly positive, although consistent with zero.

This simple argument disfavors the tachyon explanation of the OPERA data. The same conclusions
were drawn in refs. \cite{Lipari,Caccia} (appeared when this paper was completed), carrying out a detailed numerical analysis 
and including in the fit not only the recent OPERA data but also the Fermilab data, which do not display any energy dependence too. 

One could however still question this conclusion, since the energies  $\bar E_1=13.9$ GeV and $\bar E_2=42.9$ GeV quoted by OPERA
are mean ones and if we consider the $3\sigma$ range associated to $\delta t_2 -\delta t_1$ we find the interval $[-65, 93]$ ns.

\subsection{Tachyon mass range from OPERA}

Suppose that we close an eye on the energy dependence and we stick to the interpretation of the OPERA early arrival time in terms of a tachyonic  muon neutrino. 
As we are going to discuss, arguments based solely on kinematics allow to obtain an indication for the value of the tachyonic muon neutrino mass. 

\begin{figure}[t!]
\begin{center}
\includegraphics[width=10cm]{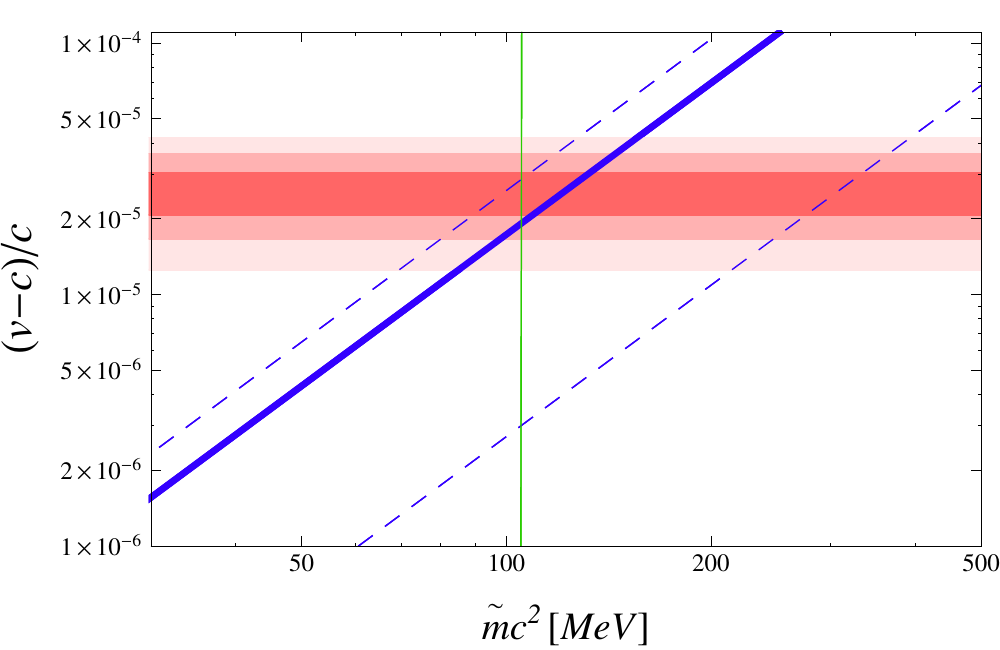} 
\end{center}\vspace*{-0.5cm} 
\caption{Relative speed of the tachyon with respect to light as a function of the tachyon mass and for selected values
of the its energy: from left to right $E=13.9,17.0,42.9$ GeV. The shaded regions display the OPERA measurement, with $1,2,3 \sigma$ error bands
(statistical and systematic errors are summed in quadrature).
The vertical line marks the value of the muon mass, $m_\mu c^2\approx 105$ MeV.}
\label{fig-dvsv}
\end{figure}

In terms of the muon neutrino energy $E$ and the muon neutrino velocity $v$, 
the tachyonic muon neutrino mass $\tilde m $ is simply given by (see eq.(\ref{deltav})):
\beq
{\tilde m} c^2 \approx\sqrt{ 2  \frac{v-c}{c}} E \,\,.
\eeq
Since OPERA deals with neutrinos with mean energy $\langle E \rangle = 17$ GeV and observes $ (v-c)/c= (2.48 \pm 0.28 \pm 0.30) \times 10^{-5}$, 
the corresponding tachyonic mass value is $\tilde m c^2=(110 -130)$ MeV at $1\sigma$, and $(85-146)$ MeV at $3\sigma$ 
(statistical and systematic errors are summed in quadrature)\footnote{Clearly this value is just a rough estimate, due to the significant energy spread 
of the neutrino beam, but we said that we ignore this as it would also cause an energy dependence of the early arrival time.} . 
This is also graphically shown in fig.\ref{fig-dvsv}. 

This range of values is close to the muon mass, $m_\mu c^2\approx 105$ MeV, and it would be fascinating to postulate
that the muon neutrino mass is just the charged muon one, upon a rotation of $\pi/2$ in the complex plane.
However, these conjectures are going to be severely challenged by other experimental data. 

\subsection{Production from pion}

Indeed, first of all we must wonder whether a tachyonic muon neutrino with such a large mass could ever be produced in pion decay: $\pi \rightarrow \mu \nu_\mu$.
The OPERA muon neutrinos are in fact obtained from pions decaying in flight in a $1$ km long vacuum tunnel\footnote{A proton beam of $400$ GeV$/c$ from SPS
hits a graphite target producing pions which are focused by two magnetic horns and directed towards the tunnel.}. 

Let us focus on the kinematics of pion decay at rest. Clearly, momentum conservation requires the muon and the neutrino to be produced back to back and with the same momentum $p$. Energy conservation requires in addition:  
\beq
m_\pi c^2 = \sqrt{(p c)^2+(m_\mu c^2)^2} + \sqrt{(p c)^2 - (\tilde m c^2)^2} \,\,,
\eeq
where $m_\pi, m_\mu $ stand for the pion and muon masses respectively.
As $\tilde m$ reaches its maximum allowed value when the muon neutrino has null energy ($p=\tilde m c$),
one derives an upper bound $\tilde m \le \sqrt{m_\pi^2 -m_\mu^2} \approx 92$ MeV$/c^2$. 
This limit is marginally compatible with the tachyonic neutrino mass range derived before from OPERA results, see fig.\ref{fig-dvsv}.

\begin{figure}[t!]
\begin{center}
\includegraphics[width=10cm]{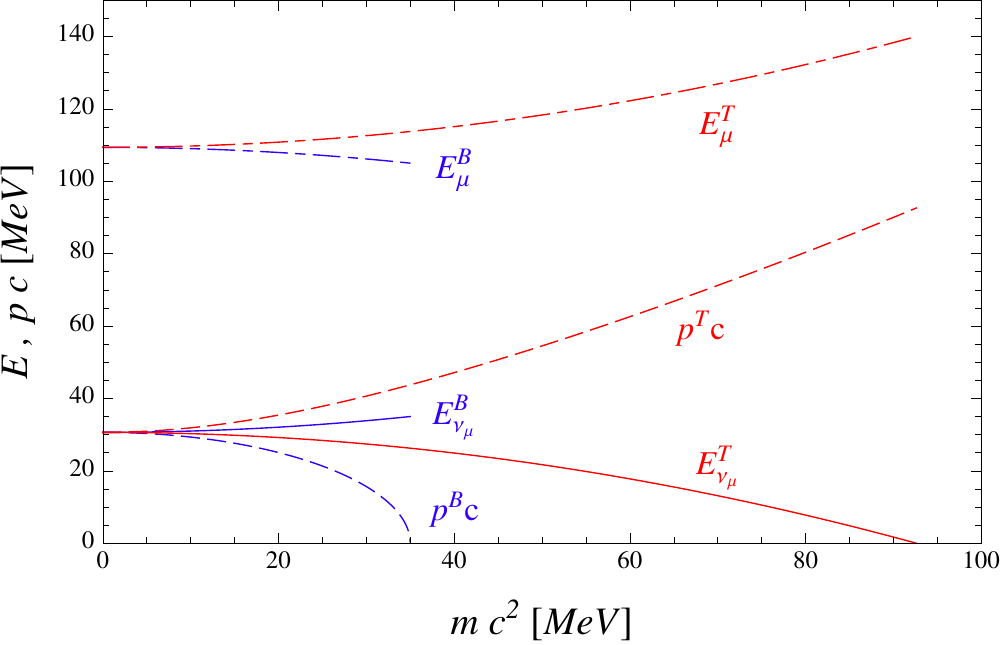} 
\end{center}\vspace*{-0.5cm} 
\caption{Kinematics for $\pi \rightarrow \mu \nu_\mu$ assuming that the muon neutrino is a tachyon (T) or a bradyon (B). The muon neutrino energy $E_{\nu_\mu}$ (solid), 
the muon energy $E_\mu$ (dot-dashed) and the associated momentum (dashed) are shown as a function of the muon neutrino mass.}
\label{fig-pi}
\end{figure}

To better support this conclusion, in fig. \ref{fig-pi} we display the $\mu$ and $\nu_\mu$ energies and momenta as a function of the muon neutrino mass, 
assuming the latter to be a tachyon (T) or a standard subluminal particle, also called bradyon  (B).
Deviations from the well tested values of the muon energy and momentum allow only a small tachyonic mass, say smaller than about $10$ MeV$/c^2$.
When inserted in eq. (\ref{deltav}) and keeping the muon neutrino energy $E = 17$ GeV, the bound $\tilde m \le10$ MeV$/c^2$ would imply $(v-c)/c \le 1.7 \times 10^{-7}$, 
which corresponds to an early arrival time at OPERA of $\delta t \le 0.6$ ns. This is at least two orders of magnitude below the observed value.

We can rephrase all this also in another way: To end up with $(v-c)/c \approx 2 \times 10^{-5}$ while keeping $\tilde m =10$ MeV$/c^2$, a muon neutrino beam 
energy $E \approx 1.4$ GeV would have been necessary. The latter value seems to be definitely too small with respect to the reconstructed muon neutrino energy.
We conclude that {\it the tachyonic explanation of the early arrival times of muon neutrinos at OPERA is ruled out}.

We also note that a kinematical analysis of muon decay ($\mu \rightarrow e  \bar \nu_e  \nu_{\mu}$)
using the tachyonic mass range suggested by OPERA would produce serious difficulties.
For simplicity, consider that  $\nu_{\mu}$ has a tachyonic mass $\tilde m$ while $e$ and $\bar \nu_e$ are massless.  
Then,  in the corner of the phase space where $E_{\nu_{\mu}}= 0$ (and consequently $p_{\nu_{\mu}} = \tilde m c$), 
the electron energy can be as high as $E_e \le m_{\mu}c^2/2 (1 + \tilde m/m_{\mu})$
(intuitively, $e$ and $\bar \nu_e$ have to balance the large momentum of the tachyon, 
and they have a lot of energy to do so, since the tachyon energy is zero). 
Even more significantly, one can see that for every value of the allowed energy range for the tachyon, $0 \le E_{\nu_{\mu}} \le m_{\mu} c^2/ 2 (1 - \tilde m^2 /m_{\mu}^2)$, 
the maximum value of $E_e$ is larger than $ m_{\mu}c^2 / 2 (1 + \tilde m^2 / m_{\mu}^2)$. 
In conclusion, values of $\tilde m\ge 10$ MeV$/c^2$ would be immediately detectable in the electron spectrum, whose endpoint would be much larger that $m_{\mu}/ 2$.
Once again, this argument rules out the hypothesis that the OPERA muon neutrino is a tachyon.

\subsection{Supernova SN1987a requires an $\lesssim$eV electron-neutrino }

Let suppose that we close another eye on the problems associated with the production of a beam of tachyonic muon neutrinos with $100$ MeV mass
and persevere on this road. Can we agree with the SN1987a data? As we are going to discuss this is possible.

The SN1987a is $L=1.68 \times 10^5$ ly far from the Earth and exploded releasing a huge neutrino signal, with typical energies $10-20$ MeV,
which allowed the first direct detection of astrophysical neutrinos. 
All neutrino flavors were emitted but Kamiokande-II, IMB and Baksan were designed to detect mainly electron anti-neutrinos.  
The signal lasted about $10$ s and the photons also arrived within a few hours. See for instance ref. \cite{Pagliaroli:2008ur} for a recent review
and a list of references.

The advance (delay) of a tachyonic (bradyonic) anti-neutrino with respect to light is $\delta T= T |v-c|/c$, where $T=L/c$ is the time associated 
to the photon trip from SN1987a to the Earth\footnote{The SN exploded when the Earth was in the quaternary period and on such timescales  effects due to the 
expansion of  the universe can be neglected.}. The fact that photons and electron anti-neutrinos arrived without few hours implies that $\delta T/T =|v-c|/c \sim 10^{-9}$,
which in turn translates into an upper bound for the (bradyonic or tachyonic) mass of the electron anti-neutrino of about $1$ keV.
 
Electron anti-neutrinos arrived with a time spread $\Delta T \lesssim 10$ s, as indicated by observations. This poses a much tighter limit on their (tachyonic or bradyonic)
mass $m_{\bar \nu_e}$ than the one just discussed. 
The time spread $\Delta T=|T_2-T_1|$ of two neutrinos with energies $E_1$ and $E_2$ (with $E_1\le E_2$) is
\beq
\frac{\Delta T}{T} =  \frac{ m_{\bar \nu_e}^2}{2 E_1^2} \left( 1- \frac{E_1^2}{E_2^2}\right)\,\,.
\label{spread}
\eeq
For the numerical values mentioned above, one obtains $m_{\bar \nu_e} \lesssim 40$ eV$/c^2$.
This limit, applies however to the electron anti-neutrino\footnote{We recall that an even stronger bound on the mass of a tachyonic electron neutrino
follows from tritium beta decay experiments \cite{Ciborowski:1998kc}, which set a limit of few eV$/c^2$.}.

The SN emits all neutrino flavors. Let suppose that it emits also a $100$ MeV tachyonic muon neutrino:
its advance with respect to light would be of about $\delta t \approx 4 yr$, but with an enormous spread as can it can be realized by considering eq.(\ref{spread})
in the case of a particle with mass bigger than its energy. These neutrinos would have certainly escaped detection.

\subsection{Oscillations: game over}

According to the picture emerged so far, the ratio between the tachyonic mass of the muon neutrino suggested by OPERA and the mass of the electron
anti-neutrino suggested by SN1987a would be as large as $10^5$.
In principle, the formalism of neutrino oscillation in the tachyonic case is the same as for an ordinary neutrino \cite{Caban:2003rb}, 
but it appears difficult to come to pact  with the robust informations coming from neutrino oscillation experiments.
Indeed, these experiments put stringent bounds on the difference of neutrino masses squared: 
$|\Delta m^2_{32}| \approx 2.4 \times 10^{-3}$ eV$^2$ and  $\Delta m^2_{21} \approx 7.6\times 10^{-5}$ eV$^2$ \cite{pdg}. 
Even though the analysis of the experimental data should be redone since the fluxes at the production in the
Sun for electron anti-neutrinos and in the atmosphere for muon neutrinos would change, 
it seems hopeless to find an agreement with experimental data on oscillations.

\section{Muon neutrino \'a  la Coleman-Glashow}
\label{secCG}

As an alternative scenario, consider now two CG neutrino mass eigenstates with masses $m_1$ and $m_2$ not larger than ${\cal O}( $eV$)/c^2$,
with different limit speeds $c_1$ and $c_2$  \cite{Coleman:1997xq, Coleman:1998ti}. 
One may infer $|c_1-c|/c \lesssim 10^{-9} $ from SN1987a as discussed in the previous section, while $(c_2-c)/c\approx 2.5\times10^{-5}$ as suggested by OPERA. 

Let us suppose that $\nu_2$ has a significant mixing with the muon neutrino of the OPERA beam, and that $\nu_1$ mixes significantly with the electron neutrino.
We now revisit for this CG superluminal neutrino model the same issues discussed for the tachyon.

First of all, the early arrival time of the muon neutrino beam is energy independent, since now $c_2$ is a constant (already chosen to reproduce the results from OPERA)
and we assumed $m_2c_2^2  \le {\cal O}($eV$)$; with these assumptions the standard kinematics used for event reconstruction at OPERA need not be modified. 

At first glance, one could envisage no problem for the production of such CG muon neutrino from pion decay. 
However, a careful kinematical analysis reveals that the situation should be considered in more details.

We first assume that $c_\pi = c_\mu= c$. For the sake of the comparison with OPERA, 
the relevant configuration is the one in which the CG muon neutrino and the muon have negligible transverse momenta with respect to the pion momentum,
so that the CG muon neutrino is actually emitted in the Gran Sasso direction. In this case, there is an upper bound for the CG muon neutrino energy 
(see also \cite{Cowsik:2011wv, Bi:2011nd}):
\beq
E_\nu \le \frac{(m^2_\pi-m^2_\mu)c^4}{2 E_\pi } \frac{c}{c_2-c}  \approx 3 \,{\rm GeV}\,\,,
\eeq
where the numerical  value is obtained by considering $E_\pi \sim 60$ GeV, together with the OPERA result $(c_2-c)/c\approx 2.5\times10^{-5}$. 
Clearly, this bound is violated by OPERA, that detects muon neutrinos with energies much larger than this value. 

A widely different scenario follows if one assumes $c_\pi = c$ and $c_\mu =c_2$. 
In this case the pion decay is forbidden unless the pion energy $E_\pi$ is smaller than a threshold energy given by
\beq
E_\pi \le \left( \frac{ (m^2_\pi-m^2_\mu)c^4}{2} \frac{c}{c_2-c} \right)^{1/2} \approx 13 \,{\rm GeV}\,\,,
\eeq
where the OPERA result for $(c_2-c)/c$ has been used. Clearly, muon neutrinos should have $E_\nu \le E_\pi$.
Again, this is in contrast with observation since the mean energy of the pions produced at CERN is about $60$ GeV and
OPERA detects neutrinos with energies up to about $80$ GeV.

This discussion shows that the observed phenomenology depends critically on the actual values of the $c_i$'s of the three particles involved in pion decay, 
whose values have been recently reviewed in \cite{Bi:2011nd}. Since at this stage a discussion of all possibilities would be inconclusive, 
we do not elaborate further on this point.

As for the SN1987a, a beam of  CG $\nu_2$ would pose no problem, because it would have simply arrived about $4$ yr in advance 
with respect to the photons and the other $\nu_1$'s. Most probably it would have escaped detection since the detectors had a lower sensitivity to muon neutrinos
(moreover Kamiokande-II started taking data only in 1985).  At variance with the tachyon case, it is important to remark that the $\nu_2$ beam
does not spread out in time but all the neutrinos arrive within a few seconds, because we assumed their mass to be smaller than eV/$c_2^2$.

A serious problem for CG neutrinos is again due to neutrino oscillations, as can be shown by using the formalism of refs.\cite{Coleman:1997xq, Coleman:1998ti}.
The two CG neutrino eigenstates travel at different speeds and this affects the neutrino oscillation probability similarly to a difference in mass:
\beq
P(\nu_\ell \rightarrow \nu_\ell ) = 1 - \sin^2 2\theta  \, \sin^2\left(\frac{R}{\hbar \bar c} \left(\frac{(m_2^2-m_1^2) {\bar c}^4}{4 E} + \frac{\delta c}{\bar c} \frac{E}{2}\right) \right) \,\,,
\label{osc}
\eeq
where $\theta$ is the mixing angle, $R$ is the distance from source to detector, $\bar c= (c_1 + c_2)/2$, $\delta c=c_2-c_1$ 
and $E$ is the neutrino energy, typically in the range of a few MeV for reactor and solar experiments. For numerical estimates, it is perfectly safe to replace $\bar c$ with $c$.
Oscillation experiments (see for instance \cite{:2008ee}) indicate a value for $m_2^2-m_1^2\approx 10^{-4}$ eV$^2$/$c^4$.  
This translates in a sensitivity to $\delta c/c$ of about $10^{-18}$, much smaller than what would be needed to explain the OPERA data.
This can be seen as follows: the experimentally tolerated oscillation frequency is the one of the first term of the $\sin^2$ argument in eq.(\ref{osc})
with $(m_2^2-m^2_1) c^4 \sim 10^{-4} $ eV$^2$ and $E \sim$ MeV. 
A comparable frequency would result from the second term of the argument of $\sin^2$ only if $ \delta c/c \sim 10^{-18}$.
Also due to the different energy dependence of these two terms, it seems unlikely that a cancellation might be at work for a much larger value of $\delta c/c \sim 10^{-5}$
in the energy range probed by oscillation experiments.
A similar analysis was done in refs.\cite{Coleman:1997xq, Coleman:1998ti}, suggesting an even tighter limit $\delta c/c \sim 6 \times 10^{-22}$. 

In conclusion, also CG superluminal neutrinos seem not to provide a fully satisfactory explanation of the OPERA results.

\section{Conclusions}
\label{concl}

 The evidence for muonic neutrinos traveling slightly faster than light in vacuum as reported by the OPERA Collaboration,
motivated us to explore two possible interpretations of the data:  the hypothesis that the muon neutrino is a tachyon or that it is a Coleman-Glashow neutrino. 

We demonstrated that the tachyonic interpretation is hardly reconciled with the energy independence of the OPERA data,
as shown also by \cite{Lipari,Caccia}.  The real problem that we  point out here is that it would be impossible to produce a $100$ MeV tachyon from pion decay.
The data associated with SN1987a can be interpreted by assuming an eV electron anti-neutrino. 
This picture however clashes with what is known concerning neutrino oscillations. 

A Coleman-Glashow superluminal neutrino beam would presumably face problems with kinematics, but it is difficult to asses details here because
the kinematic bounds strongly depend on the actual values of the limiting speeds of the particles involved in pion decay.
A CG neutrino could be easily reconciled with SN1987a data but, on the other hand, it would be not possible to reconcile the model with neutrino oscillations.  


In conclusion, the picture emerging from combining OPERA with other experimental data is that
neutrinos should not obey a tachyon type nor a Coleman-Glashow type dispersion relation,
but rather a dispersion relation with a very peculiar energy dependence \cite{Caccia, Giudice:2011mm}.
Even in this case, a serious problem could be represented by energy losses due to electron positron pair production \cite{Cohen:2011hx}.


\section*{Acknowledgements}

We thank Barbara Ricci for useful discussions. We also thank Mofazzal Azam, Cesare Malag\`u, Fabio Mantovani and Stefania Vecchi for nice comments.




\begin{thebibliography}{99}

\bibitem{opera}
Opera Collaboration, {\it Measurement of the neutrino velocity with the OPERA detector in the CNGS beam},
arXiv:1109.4897

\bibitem{Adamson:2007zzb}
  P.~Adamson {\it et al.} [ MINOS Collaboration ],
  {\it Measurement of neutrino velocity with the MINOS detectors and NuMI neutrino beam},
  Phys.\ Rev.\  {\bf D76 } (2007)  072005.
  [arXiv:0706.0437 [hep-ex]].
  
\bibitem{Kalbfleisch:1979rm}
  G.~R.~Kalbfleisch, N.~Baggett, E.~C.~Fowler, J.~Alspector,
 {\it Experimental Comparison Of Neutrino, Anti-neutrino, And Muon Velocities},
  Phys.\ Rev.\ Lett.\  {\bf 43 } (1979)  1361.
  

\bibitem{Chodos:1984cy}
  A.~Chodos, A.~I.~Hauser, V.~A.~Kostelecky,
  {\it The Neutrino As A Tachyon},
  Phys.\ Lett.\  {\bf B150 } (1985)  431.
  
\bibitem{Recami:1984xh}
  E.~Recami,
 {\it Classical Tachyons And Possible Applications: A Review},
  Riv.\ Nuovo Cim.\  {\bf 9N6 } (1986)  1-178.
  
  
\bibitem{Coleman:1997xq}
  S.~R.~Coleman, S.~L.~Glashow,
 {\it Cosmic ray and neutrino tests of special relativity},
  Phys.\ Lett.\  {\bf B405 } (1997)  249-252.
  [hep-ph/9703240].
  
\bibitem{Coleman:1998ti}
  S.~R.~Coleman, S.~L.~Glashow,
 {\it High-energy tests of Lorentz invariance},
  Phys.\ Rev.\  {\bf D59 } (1999)  116008.
  [hep-ph/9812418].  
  
  
 \bibitem{Lipari}
 G. Amelino-Camelia, G. Gubitosi, N. Loret, F. Mercati, G. Rosati, P. Lipari,
{\it OPERA-reassessing data on the energy dependence of the speed of neutrinos},
   arXiv:1109.5172 
  
 \bibitem{Caccia}
 G. Cacciapaglia, A. Deandrea, L. Panizzi
{\it Superluminal neutrinos in long baseline experiments and SN1987a},
  arXiv:1109.4980 
  
  
\bibitem{Pagliaroli:2008ur}
  G.~Pagliaroli, F.~Vissani, M.~L.~Costantini, A.~Ianni,
 {\it Improved analysis of SN1987A antineutrino events},
  Astropart.\ Phys.\  {\bf 31 } (2009)  163-176.
  [arXiv:0810.0466 [astro-ph]].
  
\bibitem{Ciborowski:1998kc}
  J.~Ciborowski, J.~Rembielinski,
  {\it Tritium decay and the hypothesis of tachyonic neutrinos},
  Eur.\ Phys.\ J.\  {\bf C8 } (1999)  157-161.
  [arXiv:hep-ph/9810355 [hep-ph]].  
     
\bibitem{Caban:2003rb}
  P.~Caban, J.~Rembielinski, K.~A.~Smolinski, Z.~Walczak,
 {\it Oscillations do not distinguish between massive and tachyonic neutrinos},
  Found.\ Phys.\ Lett.\  {\bf 19 } (2006)  619-623.
  [hep-ph/0304221].

 \bibitem{pdg}
 K. Nakamura et al. (Particle Data Group), J. Phys. G 37, 075021 (2010) 
and 2011 partial update for the 2012 edition. 

\bibitem{Cowsik:2011wv}
  R.~Cowsik, S.~Nussinov, U.~Sarkar,
  {\it Superluminal Neutrinos at OPERA Confront Pion Decay Kinematics},
   [arXiv:1110.0241 [hep-ph]].

\bibitem{Bi:2011nd}
  X.~-J.~Bi, P.~-F.~Yin, Z.~-H.~Yu, Q.~Yuan,
  {\it Constraints and tests of the OPERA superluminal neutrinos},
   [arXiv:1109.6667 [hep-ph]].

\bibitem{:2008ee}
  S.~Abe {\it et al.} [ KamLAND Collaboration ],
  {\it Precision Measurement of Neutrino Oscillation Parameters with KamLAND},
  Phys.\ Rev.\ Lett.\  {\bf 100 } (2008)  221803.
  [arXiv:0801.4589 [hep-ex]].

\bibitem{Giudice:2011mm}
  G.~F.~Giudice, S.~Sibiryakov, A.~Strumia,
 {\it Interpreting OPERA results on superluminal neutrino},
   [arXiv:1109.5682 [hep-ph]].

\bibitem{Cohen:2011hx}
  A.~G.~Cohen, S.~L.~Glashow,
  {\it New Constraints on Neutrino Velocities},
  [arXiv:1109.6562 [hep-ph]].

\end{thebibliography}
\end{document}